# Oxide tunnel junctions supporting a two-dimensional electron gas

J. D. Burton,[1*] J. P. Velev[2] and E. Y. Tsymbal[1]

[1]Department of Physics and Astronomy, Nebraska Center for Materials and Nanoscience, University of Nebraska, Lincoln, Nebraska 68588-0111, USA

[2]Department of Physics and Institute for Functional Nanomaterials, University of Puerto Rico, San Juan, PR 00931-3344, USA

The discovery of a two-dimensional electron gas (2DEG) at the interface between insulating oxides has led to a well-deserved level of excitement due to possible applications as "in-plane" all-oxide nanoelectronics. Here we expand the range of possibilities to the realm of "out-of-plane" nanoelectronics by examining such all-oxide heterostructures as barriers in tunnel junctions. As an example system we perform first-principles electronic structure and transport calculations of a tunnel junction with a $[SrTiO_3]_4/[LaO]_1/[SrTiO_3]_4$ heterostructure tunneling barrier embedded between $SrRuO_3$ electrodes. The presence of the LaO atomic layer induces the formation of a 2DEG within the tunneling barrier which acts as an extended defect perpendicular to the transport direction, providing a route for resonant tunneling. Our calculations demonstrate that the tunneling conductance in this system can be strongly enhanced compared to a pure $SrTiO_3$ barrier due to resonant tunneling, but that lattice polarization effects play a significant role in determining this behavior. In addition we find that this resonant tunneling is highly selective of the orbital symmetry of the tunneling states due to the "orbital polarization" of the 2DEG. We also discuss how the properties of the 2DEG are affected by the presence of metal electrodes.

## I. INTRODUCTION

Advances in thin-film deposition and characterization techniques have made possible the experimental realization of oxide heterostructures with atomically abrupt interfaces. The development of such heterostructures is very promising as it offers novel functionalities and device concepts. In particular the discovery of metallic conductivity at the interface between insulating oxides La$BO_3$ ($B$ = Al or Ti) and $SrTiO_3$ has induced a great amount of interest in these systems.[1,2] This (quasi) two-dimensional electron gas (2DEG) formed at the n-type $LaO/TiO_2$ interface has high carrier mobility and electron density, making it attractive for applications in nanoelectronics, e.g. as all-oxide high-mobility field-effect transistors.[3,4] It was demonstrated that at low temperatures the 2DEG could become magnetic[5] or superconducting.[6] The 2DEG was also found at $LaVO_3/SrTiO_3$ interfaces[7] and was predicted to occur at $KNbO_3/SrTiO_3$[8] and $LaAlO_3/EuO$[9] interfaces adding new functionalities to the system.

The properties of the 2DEG depend strongly on sample preparation conditions. When small oxygen pressure is used during deposition oxygen vacancies can have significant contribution to the conductivity.[10-12] However, under sufficiently high oxygen pressure the intrinsic effects dominate.[12] Intrinsically, the 2DEG is a result of the polar discontinuity effect arising from the fact that La$BO_3$ consists of alternating $(LaO)^+$ and $(BO_2)^-$ charged planes and $SrTiO_3$ consists of alternating $(SrO)^0$ and $(TiO_2)^0$ neutral planes. In semiconductors such a divergence of the electric potential caused by polar interfaces can be avoided by atomic reconstruction at the interface.[13,14] In the case of the $LaO/TiO_2$ interface, however, mixed valence states of Ti allow for electronic reconstruction. Half of an electron per two-dimensional unit cell from $LaAlO_3$ to $SrTiO_3$ is transferred to the Ti 3$d$ conduction bands, thus, reducing its valence from $Ti^{4+}$ (as in bulk $SrTiO_3$) toward $Ti^{3+}$ and making the interface conducting.[15] A charge transfer to the interface also occurs if the $LaAlO_3$ layer is non-stoichiometric and terminated with the LaO monolayers on both sides (in a limiting case a monolayer of LaO replaces a monolayer of SrO in a $SrTiO_3$ (001) crystal). In this case an "extra" electron is introduced into the system due to the uncompensated ionic charge on the additional $(LaO)^+$ monolayer. This electronic charge is accommodated by partially occupying conduction band states near the interface, producing a 2DEG. The electronic reconstruction mechanism is captured by first principles calculations of La$BO_3/SrTiO_3$ superlattices with stoichiometric $SrTiO_3$, based on density functional theory within the local density approximation (LDA)[16-20] and the LDA+$U$ approximation.[21-24] Calculations show the presence of $n$-type charge carriers (about ½ electron on the interface Ti-3$d$ band) at the $LaO/TiO_2$ interface.

The nature of the confinement of the 2DEG was recently addressed. The current distribution was mapped by scanning the cross-section with a conducting atomic force microscope tip, and the 2DEG was found to be localized within a few nm at the interface.[25] These findings were corroborated by angle-dependent hard X-ray photoelectron spectroscopy studies indicating that the 2DEG is indeed confined to a few unit cells at the interface.[26] Based on first principles calculations the characteristic confinement width was found to be about 1nm, determined by a mechanism identical to the metal induced gap states in semiconductors.[27]

Given that the 2DEG is confined close to the interface, a conceptually new type of tunnel junction can be conceived by employing a La$B$O$_3$/SrTiO$_3$ (or similar) layered composite as a tunneling barrier. Due to the 2DEG formed within the barrier layer parallel to the metal/insulator interface, such a tunnel junction may exhibit a resonant transmission between the electrodes. These *resonant tunneling junctions* (RTJs) may have new interesting properties and potential applications. For example, they can be used for vertical measurements to characterize properties of the 2DEG.[28] In addition, the RTJs involving magnetic electrodes could have much lower resistance compared to the conventional magnetic tunnel junctions (MTJs) and therefore may be useful for magnetic recording applications (for a review on MTJs and tunneling magnetoresistance see, e.g., ref. [29]). Moreover, experimental and theoretical data suggests that the 2DEG could be magnetic.[5,23] In this case the 2DEG is spin polarized, RTJs could be used for spin filtering and could have applications in generating spin polarized currents.

In this paper we perform first-principles electronic structure and transport calculations of all oxide heterostructure supporting a 2DEG within the barrier of a tunnel junction. As an example system we consider a RTJ with a [SrTiO$_3$]$_4$/[LaO]$_1$/[SrTiO$_3$]$_4$ heterostructure tunneling barrier embedded between SrRuO$_3$ electrodes. The presence of the LaO atomic layer induces the formation of a 2DEG within the tunneling barrier which acts as an extended defect perpendicular to the transport direction, providing a route for resonant tunneling. Our calculations demonstrate that the tunneling conductance in this system can be strongly enhanced compared to a pure SrTiO$_3$ barrier due to resonant tunneling, but that lattice polarization effects play a significant role in determining this behavior. In addition we find that this resonant tunneling is highly selective of the orbital symmetry of the tunneling states due to the "orbital polarization" of the 2DEG.

## II. STRUCTURES AND METHODS

We consider three related tunnel junction structures as shown in Fig. 1. First, as a reference system, we consider a SrRuO$_3$/[SrTiO$_3$]$_8$/SrRuO$_3$ tunnel junction of metal SrRuO$_3$ electrodes separated by 8½ unit-cells of SrTiO$_3$ stacked along the [001] direction of the conventional perovskite cell. We designate this structure as TJ$_0$. Then, we construct the SrRuO$_3$/[SrTiO$_3$]$_4$/[LaO]$_1$/[SrTiO$_3$]$_4$/SrRuO$_3$ tunnel junction by replacing the central SrO atomic layer with LaO. In this case we slightly expand the initial supercell along the $z$-direction to accommodate a larger interlayer Ti-Ti distance between the two TiO$_2$ atomic layers on either side of the LaO atomic layer. This is done to be consistent with a tetragonal distortion of $c/a = 1.029$ found from a calculation of bulk LaTiO$_3$ constrained to the in-plane lattice constant of SrTiO$_3$. Otherwise, no other relaxation is performed beyond that found for TJ$_0$. The last structure we consider, TJ$_2$, is chemically identical to TJ$_1$ but the atomic structure is fully relaxed.

We use a plane-wave pseudopotential approach as implemented in the Quantum-ESPRESSO package.[30] The exchange-correlation functional is treated in the Perdew-Burke-Ernzerhof generalized gradient approximation.[31] Atomic relaxation calculations were performed using a 6×6×1 Monkhorst-Pack grid for k-point sampling and an energy cutoff of 400 eV for the plane wave expansion. Atomic positions are converged until the Hellmann-Feynman forces on each atom became less than 20 meV/Å. Subsequent non-self-consistent density of states (DOS) calculations are performed using a 20×20×2 Monkhorst-Pack grid for k-point sampling. The in-plane lattice constant of the supercells are constrained to the calculated value for bulk cubic SrTiO$_3$, $a = 3.937$Å, to simulate epitaxial growth on a SrTiO$_3$ substrate. This theoretical lattice constant overestimates the experimental value of $a = 3.905$Å, however, using the theoretical values prevents any unphysical strain effects or tetragonal distortions in the SrTiO$_3$ barrier. The epitaxial strain on the SrRuO$_3$ induces a tetragonal distortion with $c/a = 1.050$.

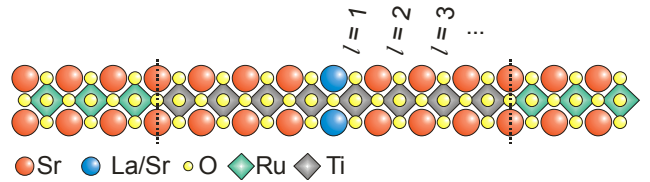

FIG. 1. (color online) Atomic structure of the supercells used in the calculations. The system has mirror symmetry about the central atomic layer, which is either SrO (TJ$_0$) or LaO (TJ$_1$ and TJ$_2$). The index $l$ denotes the $l^{th}$ $B$O$_2$ layer ($B$ = Ti or Ru) away from the central LaO/SrO layer.

Calculations are performed allowing for spin-polarization since SrRuO$_3$ is known to be a weak ferromagnet with the Curie temperature of about 160 K. However most effects we describe in this paper are equally applicable to a non-spin-polarized case since our focus is on describing the tunneling through a non-spin-polarized 2DEG in trapped in a barrier.

The results of atomic relaxations are shown in Fig. 2 where we plot, layer-by-layer, the polar displacements perpendicular to the plane of the junction between the metal cation (Sr, La, Ti or Ru) with respect to its oxygen neighbors in the same (001) atomic plane. For the TJ$_0$ structure with the pure SrTiO$_3$ barrier we do not find any large polar displacements. However, for the fully relaxed junction with the LaO layer, TJ$_2$, we do indeed find substantial polar displacements in the layers composing the SrTiO$_3$ barrier. These displacements are more-or-less uniform in magnitude but opposite in sign on either side of the central LaO layer. This corresponds to the development of a peculiar polarization formation that is reminiscent of an abrupt, tail-to-tail domain wall between ferroelectric-like domains. This is in contrast to previous studies of similar

SrTiO$_3$/LaTiO$_3$ heterostructures without metal electrodes where the polar distortions in the SrTiO$_3$ were found to fall off away from the LaO layer.[21,32,33] The origin of the lattice polarization effects shown in Fig. 2 is discussed in section IV-B.

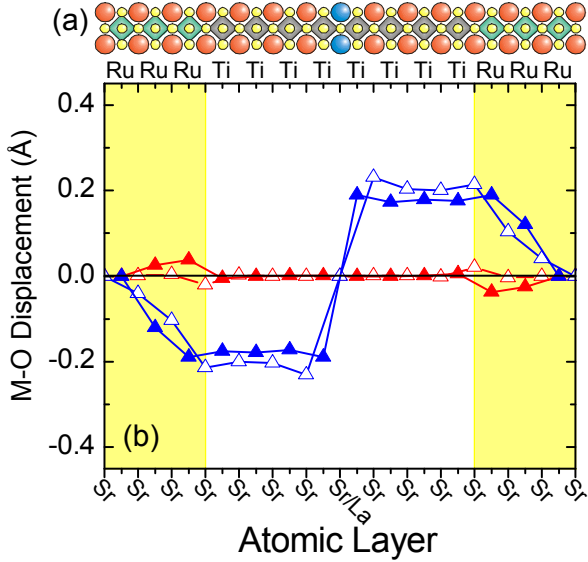

FIG. 2. Cation-anion polar displacements across the tunnel junctions. The red curves are for a pure SrTiO$_3$ barrier (TJ$_0$) as well as the assumed structure of TJ$_1$. The blue curves are the fully relaxed system where the central atomic layer is LaO (TJ$_2$). The solid symbols are $B$-O$_2$ displacements ($B$ = Ru or Ti) and the open symbols are $A$-O displacements ($A$ = Sr or La).

## III. TRANSPORT CALCULATIONS

The tunneling properties of these junctions are calculated using a general scattering formalism adapted to ultra-soft pseudopotentials[34,35] and fully implemented in the Quantum-ESPRESSO package. The structures in Fig. 1 are considered as the scattering region, ideally attached on both sides to semi-infinite SrRuO$_3$ leads. Three unit cells of SrRuO$_3$ on each side of the barrier are found to be sufficient to reproduce a bulk-like potential on both sides of the scattering region. Transmission and reflection matrices are then obtained by matching the wave functions in the scattering region to appropriate linear combinations of the Bloch states in the left and right leads. In the zero-bias limit conductance is evaluated by calculating the electron transmission for states at the Fermi level. The conductance per unit cell area is given by the Landauer-Büttiker formula

$$G = \frac{e^2}{h}\sum_{\sigma \mathbf{k}_\|} T_\sigma(\mathbf{k}_\|), \quad (1)$$

where $T_\sigma(\mathbf{k}_\|)$ is the transmission probability of the electron with spin $\sigma$ at the Fermi energy. Since our system has perfect periodicity in the plane perpendicular to the transport direction the Bloch wave vector $\mathbf{k}_\| = (k_x, k_y)$ is preserved in tunneling. The total tunneling conductance is found by integration over the 2D Brillouin zone using a uniform $100 \times 100$ $\mathbf{k}_\|$ mesh.

**Table I.** The spin-resolved zero-bias conductance $G$, in units of $e^2/h$, of the three tunnel junctions. The right-most column shows the spin polarization (SP) of the tunneling current.

|  | Majority | Minority | Total | SP |
|---|---|---|---|---|
| TJ$_0$ | $1.29\times10^{-8}$ | $0.58\times10^{-8}$ | $1.87\times10^{-8}$ | 38% |
| TJ$_1$ | $0.40\times10^{-3}$ | $1.24\times10^{-3}$ | $1.64\times10^{-3}$ | -51% |
| TJ$_2$ | $0.08\times10^{-8}$ | $0.25\times10^{-8}$ | $0.33\times10^{-8}$ | -54% |

In Fig. 3 we plot the calculated $\mathbf{k}_\|$-resolved transmission distributions over the entire 2-D Brillouin zone for the three tunnel junctions. For the pure SrTiO$_3$ barrier (TJ$_0$) we find that for both spin-channels the largest contributions to the total transmission occur along a cross pattern centered at $\mathbf{k}_\| = 0$, i.e. at the $\bar{\Gamma}$ point. This arises due to the $\mathbf{k}_\|$-dependence of the lowest tunneling decay rate of SrTiO$_3$, as determined by the complex band structure within the band gap, consistent with previous calculations of SrTiO$_3$[27,36] and the isovalent titanate BaTiO$_3$.[37,38] The total spin-resolved conductance is listed in Table I, along with the spin polarization SP, defined as

$$SP = \frac{T_\uparrow - T_\downarrow}{T_\uparrow + T_\downarrow} \times 100\%. \quad (2)$$

There is a striking difference in the transmission distribution, however, when the central SrO layer is replaced with LaO (TJ$_1$). In Fig. 3 we see that transmission for both spin-channels is dominated by a pattern of narrow contours throughout the 2D Brillouin zone. These correspond to resonant tunneling through the junction due to the presence of a 2DEG confined in the middle of the barrier, which we discuss below. In addition, we find a significant increase in the minority-spin transmission even away from the narrow contours. Overall, the conductance of TJ$_1$ is enhanced by a factor of $10^5$ as compared to TJ$_0$ (see Table I). Allowing full relaxation after the insertion of the LaO layer (TJ$_2$), we see that the resonant features present before relaxation (TJ$_1$) disappear and, in fact, the total conductance is reduced even below that found for pure SrTiO$_3$ junction, TJ$_0$. The origins of these behaviors are discussed in the next sections.

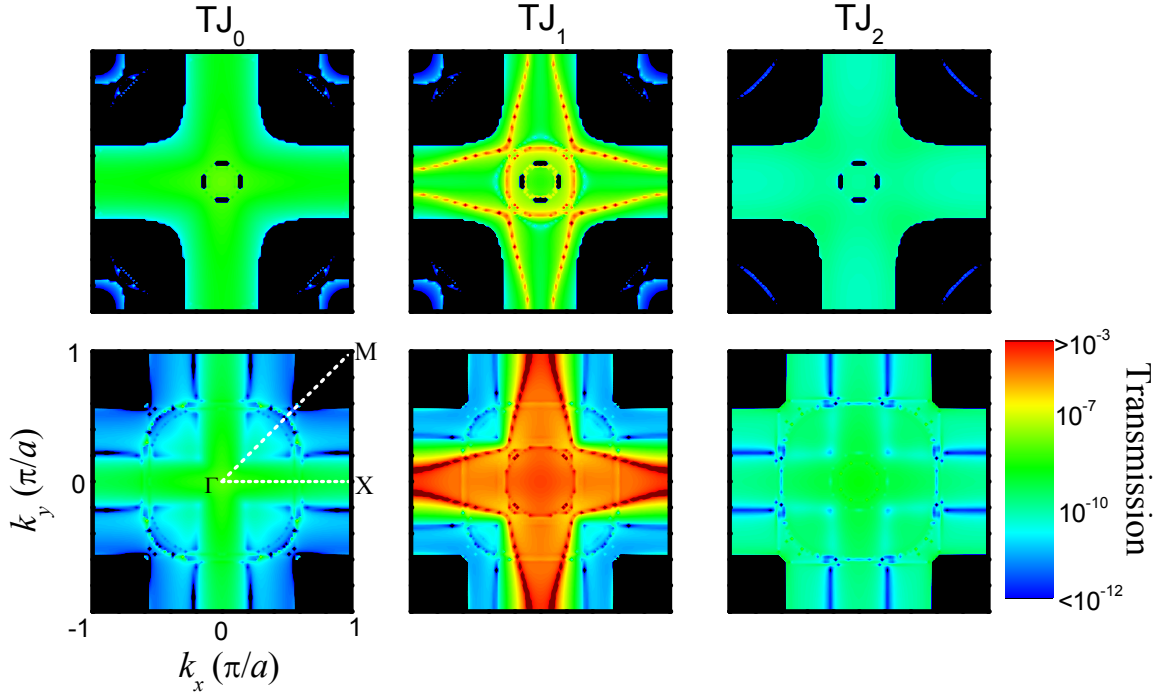

FIG. 3. The $\mathbf{k}_{\|}$-resolved distribution of the transmission through the $SrRuO_3/(Sr,La)TiO_3/SrRuO_3$ tunnel junctions in the 2D Brillouin zone. The top row is for majority spin states and the bottom row is for minority spin states. The first column corresponds to $TJ_0$, the second column to $TJ_1$ and the third column to $TJ_2$. Note that the color scale is logarithmic.

## IV. DISCUSSION
### A. Electronic structure of the junctions

In a generic cubic perovskite crystal, $ABO_3$, the $d$ states of the transition metal $B$ atom are split by the octahedral crystal field produced by the O cage into $t_{2g}$ states ($d_{zx}$, $d_{zy}$, $d_{xy}$) and $e_g$ states ($d_{z^2}$ and $d_{x^2-y^2}$), with the $t_{2g}$ levels lying lower in energy. In the crystal these states are broadened into bands. In the $d^0$ system of $SrTiO_3$ these states are completely empty and therefore constitute the conduction bands, whereas the valence bands are mainly formed by the O-$2p$ states. In $SrRuO_3$ the partially occupied Ru-$d$ states reside around the Fermi level and are mainly responsible for the metallic behavior. In the tunnel junction systems we consider here, those states that derive from bands with out-of-plane orbital character, i.e. $d_{z^2}$, $d_{zx}$ and $d_{yz}$, dominate the tunneling properties over those bands that derive from orbitals with in-plane orbital character, i.e. $d_{xy}$ and $d_{x^2-y^2}$. This is because band dispersion is determined by the overlap with nearest neighbors, and therefore only those orbitals with out-of-plane orbital character will have substantial overlap along the [001] direction.

Keeping this in mind, we examine the layer and orbital projected local density of states (LDOS) for the three junctions. Fig. 4 shows the LDOS on the $BO_2$ layers only, since the $A$-site projected LDOS is negligible around the Fermi level. We only plot for the layers on one side of the central $AO$ layer; the LDOS on the other side is identical owing to the mirror symmetry (see Fig. 1). In addition we only plot the majority-spin LDOS for the $TiO_2$ layers in the barrier. The minority-spin LDOS is not noticeably different even for the layer nearest the spin-polarized $SrRuO_3$ electrode.

For the case of a pure $SrTiO_3$ barrier, $TJ_0$, we find the characteristic electronic structure of a standard tunnel junction. The Fermi level lies within the band gap of the $SrTiO_3$ which means that conduction through this barrier is carried through the evanescent states, i.e. tunneling.[39] The position of the conduction band minimum (CBM) and valence band maximum (VBM) are constant across the entire barrier indicating that there are no net macroscopic electrostatic fields present in the barrier. The states near the CBM derive from the Ti-$t_{2g}$ orbitals. Notice that the $d_{zx}$ and $d_{zy}$ projected LDOS, which are identical due to the four-fold symmetry, and the $d_{xy}$ LDOS are aligned in energy due to the near perfect octahedral environment of the Ti sites.

In the case where the central atomic layer is replaced by LaO ($TJ_1$) we find a substantial difference in the electronic structure across the junction. First, the positions of the CBM and VBM are not constant across the barrier, indicating the presence of the net macroscopic electrostatic potential due to the substitution of the $Sr^{2+}$ ions with $La^{3+}$ on the central layer. In particular the CBM on layers $l = 1$ and 2 are pulled down to the extent that the conduction bands become partially occupied. These "extra" electrons that populate the conduction band states are due to the substitution of divalent Sr in the central layer with trivalent La, representing a form of n-type doping, giving rise to the oxide 2DEG (see also ref. [27]). Notice also that the Fermi level clearly lies within the gap on layers $l = 3$ and 4 indicating that this 2DEG is indeed insulated from the metal electrodes by tunneling barriers. Such a structure can be considered as a double barrier tunnel junction, where the central "potential well" consists of the 2DEG.

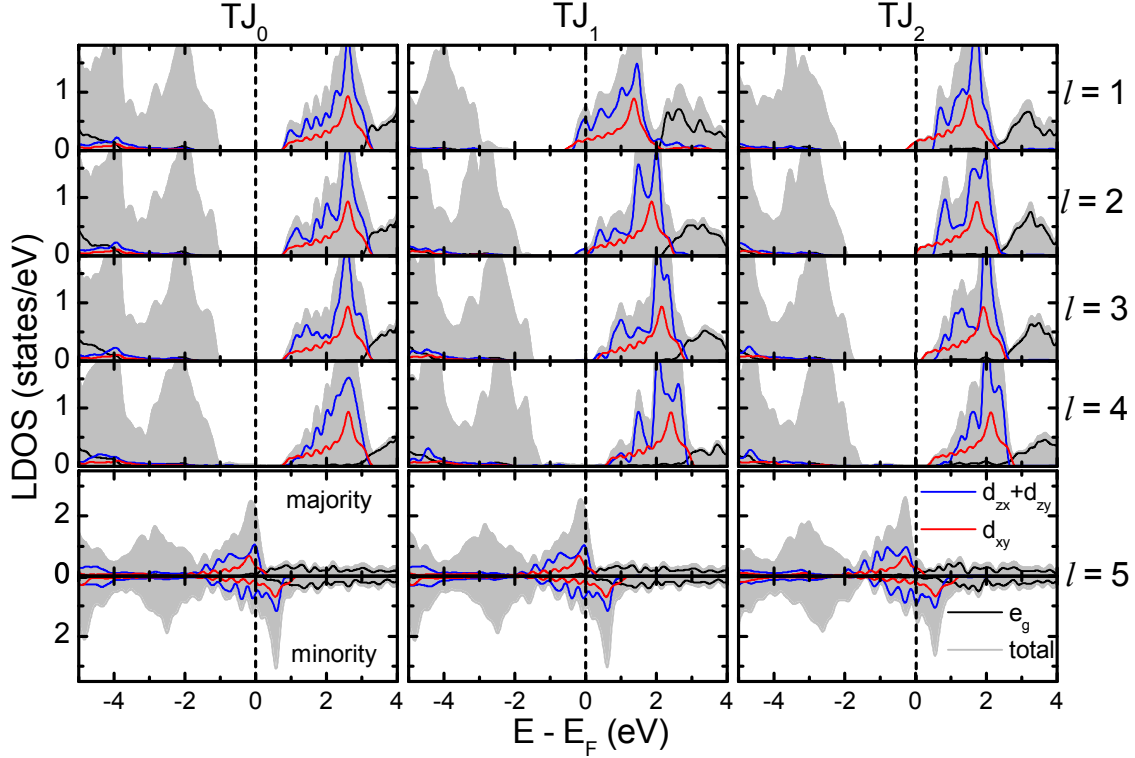

FIG. 4. The local density of states (LDOS) projected onto the $BO_2$ layers ($B$ = Ti for $l \leq 4$ and $B$ = Ru for $l = 5$) in the $SrRuO_3/(Sr,La)TiO_3/SrRuO_3$ tunnel junctions. In each subplot the grey shaded curve is the sum of the projected LDOS for all atoms (including oxygen atoms) in the $l^{th}$ $BO_2$ layer away from the central $AO$ atomic layer (see Fig. 1). The non-shaded curves are projections onto the $d$-states of the transition metal $B$ site. The black curve is the sum over the $e_g$ orbitals ($d_{z^2}$ and $d_{x^2-y^2}$), the blue curve is the sum of $d_{zx}$ and $d_{zy}$ orbitals, and the red curve is for the $d_{xy}$ orbital. For the $TiO_2$ layers ($l \leq 4$) only the majority-spin LDOS is plotted. The position of the Fermi level is indicated by the vertical dashed lines.

To see that the narrow contours that are observed in Fig. 3 in fact comes from resonant tunneling through this unconventional RTJ, we calculated the 2-D Fermi "surface" of an unrelaxed $[SrTiO_3]_{7.5}/[LaO]_1$ superlattice (i.e. no electrodes), which we plot in Fig. 5. The calculated Fermi surface matches perfectly with most of the resonant features seen for $TJ_1$ plotted Fig. 3 and re-plotted in the background of Fig 5. Notice, however, that not all of the $\mathbf{k}_\parallel$-points in Fig. 5 show up resonantly in the tunneling distribution of $TJ_1$. To understand this behavior we also calculated the orbital character of the LDOS along the Fermi contours, indicated by the point-color in Fig. 5. Just as we found in Fig 4, the states at the Fermi level are dominated by the $t_{2g}$ states ($d_{zx}, d_{zy}, d_{xy}$) on the Ti site nearest the LaO layer. The $\mathbf{k}_\parallel$-points with strong out-of-plane $d_{zx}+d_{zy}$ orbital character (blue) lead to resonant tunneling, whereas states with purely in-plane $d_{xy}$ character (red) do not. Those $\mathbf{k}_\parallel$-points with intermediate character show up in the tunneling distribution but are less pronounced. Thus the tunnel junction acts as a strong "orbital filter".

For $TJ_2$ we see from Fig. 4 that, while still maintaining an electronic structure reminiscent of a RTJ, the nature of the 2DEG is significantly different from that of $TJ_1$. First, comparing $TJ_1$ and $TJ_2$ we see a significant upward shift of the LDOS on the $l = 1$ and 2 layers indicating that the total occupation of conduction band states in the barrier is reduced. Upon relaxation some of the charge from the 2DEG is transferred to the electrodes, which can be seen by comparing the LDOS on the interfacial $RuO_2$ layer ($l = 5$) of $TJ_1$ and $TJ_2$. In addition to this depopulation of the conduction band states, we also see a significant splitting between the Ti $d_{xy}$ and $d_{zx}/d_{zy}$ densities of states. The origins of this are the polar displacements that develop along the transport direction as seen in Fig. 3. These displacements correspond to a shift of the Ti atom away from the center of its respective O-cage, thereby lifting the octahedral symmetry and splitting the $t_{2g}$ states into a low lying $d_{xy}$ singlet and a $d_{zx}+d_{zy}$ doublet. This splitting, coupled with the depopulation, removes all $d_{zx}$ and $d_{zy}$ character from the Fermi level, leaving only a small electron pocket with $d_{xy}$ character. Thus, the large lattice relaxations eliminate the resonant behavior of the tunnel junction.

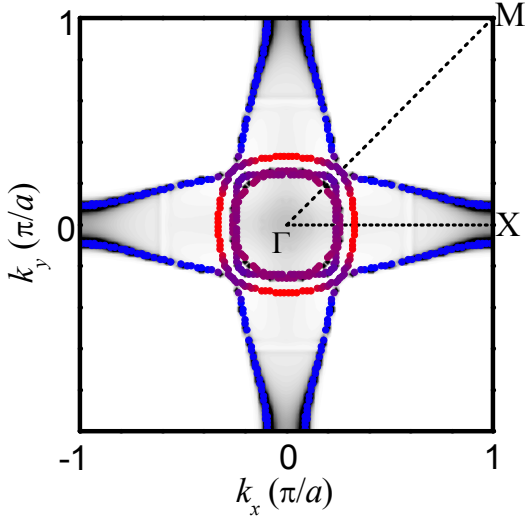

FIG. 5. Points lie along the calculated Fermi contours of an unrelaxed $[SrTiO_3]_{7.5}/[LaO]_1$ superlattice in the 2-D Brillouin zone. The point color indicates the orbital character of the $\mathbf{k}_\parallel$-resolved LDOS on the $l = 1$ Ti-d states. Red corresponds to pure $d_{xy}$ orbital character, blue corresponds to pure $d_{zx}/d_{zy}$ orbital character and the hues in between reflect an admixture of in- and out-of-plane character. The background grayscale plot is the total transmission (sum over both spin-channels) of $TJ_1$ for comparison. The transmission data is the same as that in Fig. 3, except plotted with a linear scale.

### B. Lattice polarization and charge transfer effects

Now we discuss the origins of the lattice relaxation and charge transfer effects. We consider a sheet with positive charge density $+e/a^2$ embedded at the center ($z = 0$) of a dielectric slab of thickness $2t$. This represents the $(LaO)^+$ layer inside the $SrTiO_3$ barrier. We assume for simplicity that the $SrTiO_3$ has linear dielectric constant $\varepsilon$: neglecting the non-linear dielectric response does not detract from the overall physical picture. This slab is placed between two semi-infinite ideal metal electrodes. The additional charge density due to the "extra" electron per unit cell area is $-e/a^2$. The goal of this simple model is to find the fraction, $n$, of this charge that is distributed uniformly over the positive sheet embedded in the slab, with the remainder, $(1 - n)$, distributed over the two metal/insulator interfaces. Therefore, there is a sheet of free charge density $\sigma_1 = e/a^2 - ne/a^2$ embedded within the slab, and over each metal/dielectric interface there is a free charge density $\sigma_2 = (n - 1)e/2a^2$. According to Gauss's law, these sheets of free charge produce an electric displacement

$$\mathbf{D}(z) = \frac{e(1-n)}{2a^2}\text{sgn}(z)\hat{\mathbf{z}} \qquad (3)$$

throughout the barrier, but zero outside the slab in the metal electrodes. Here $\text{sgn}(z) = +1$ for $z > 0$ and $-1$ for $z < 0$. The electric field is given by $\mathbf{E}(z) = \mathbf{D}(z)/\varepsilon$ and therefore the polarization in the slab is

$$\begin{aligned}\mathbf{P}(z) &= \mathbf{D}(z) - \varepsilon_0 \mathbf{E}(z) \\ &= \frac{e(1-n)(\varepsilon - \varepsilon_0)}{2a^2\varepsilon}\text{sgn}(z)\hat{\mathbf{z}}.\end{aligned} \qquad (4)$$

Eq. (4) implies that *if* there is charge transfer from the 2DEG to the electrodes, i.e. $n < 1$, then a tail-to-tail polarization profile develops in the dielectric, just as we see in Fig. 1. The total electrostatic energy per unit area, including the energy associated with polarizing the dielectric, is given by

$$W = \frac{1}{2}\int_{-t}^{t}\mathbf{E}(z)\cdot\mathbf{D}(z)dz = \frac{e^2(1-n)^2 t}{4\varepsilon a^4}. \qquad (5)$$

Therefore, from the point of view of electrostatics, keeping the charge of the "extra" electron in the center of the slab, i.e. $n = 1$, is energetically most favorable.

So why do we see charge transfer? The answer to this question is that there is an additional energy contribution that determines $n$: the band energy (basically the kinetic energy) of the electrons is different depending on whether the electrons reside near the $(LaO)^+$ layer or on the metal/insulator interfaces. If the band energy associated with overpopulating the surfaces of the metal electrodes is lower than that of populating the 2DEG then there will be a tendency toward transfer of charge. To obtain a qualitative estimate of this behavior we assume that the local density of states around the Fermi level is constant for both the 2DEG and at the metal surfaces. Populating the 2DEG with $n$ electrons then gives rise to band energy per unit area

$$E_{2D} \simeq \frac{n^2}{4a^2\rho_{2D}}, \qquad (6)$$

where $\rho_{2D}$ is the local density of states of the 2DEG. Here we assume the 2DEG is localized to one atomic $TiO_2$ layer on either side of the LaO. Similarly, the band energy per unit area associated with overpopulating the metal surfaces with $1 - n$ electrons is given by

$$E_M \simeq \frac{(1-n)^2}{4\delta a\rho_M}, \qquad (7)$$

where $\rho_M$ is the density of states of the metal and $\delta$ is the screening length of the metal. The energy contributions given in Eqs. (5 - 7) compete to determine the equilibrium value of $n$:

$$n = \frac{1}{1+\frac{1}{a\rho_{2D}}\left/\left(\frac{e^2 t}{\varepsilon a^3}+\frac{1}{\rho_M \delta}\right)\right.} \quad (8)$$

By examining Eq. (8) we can gain an insight into what factors contribute to the charge transfer and lattice polarization effects. Some representative parameters we estimate from the first principles calculations are $\rho_{2D} \sim 0.5$ eV$^{-1}$, $\rho_M \sim 1.0$ eV$^{-1}$ and $\delta \sim 6.2$ Å (approximately 1.5 unit cells of SrRuO$_3$). If $\varepsilon$ is very small then the electrostatic energy given in Eq. (5) is very large and therefore charge transfer will be suppressed, leaving the 2DEG largely populated. This appears to be the case for our TJ$_1$ where lattice distortions are frozen out and therefore the SrTiO$_3$ has only a very small polarizability due to the deformation of electronic degrees of freedom. In fact there is no discernable transfer of charge to the electrodes, which can be seen by comparing the LDOS on the interfacial RuO$_2$ layers ($l = 5$) in Fig. 4 for TJ$_0$ and TJ$_1$. Choosing $\varepsilon \sim 5\varepsilon_0$, corresponding to the bare electronic polarizability of SrTiO$_3$, we find $n \sim 0.95$ from Eq. (8), in good agreement with the first-principles results. On the other hand, if $\varepsilon$ is very large we see from Eq. (5) that the electrostatic cost to transfer charge to the electrodes is small. This is the case for TJ$_2$ where we allow for the large lattice polarizability of the SrTiO$_3$, leading to the large, but not complete, depletion of the 2DEG. Choosing $\varepsilon \sim 300\varepsilon_0$, corresponding to the large ionic polarizability of SrTiO$_3$, we find from Eq. (8) $n \sim 0.38$. This is in qualitative agreement with what we find for TJ$_2$, however it underestimates the amount of charge transfer because we neglect the non-linear response of the SrTiO$_3$ and also because our model neglects the ionic contribution to screening in the SrRuO$_3$, which is known to be important in polarization screening.[40] Nevertheless, as we have already seen this charge transfer and polar distortion reduces the Fermi surface of the 2DEG to a small electron pocket deriving from states that are not compatible with tunneling.

While the lattice relaxation effects remove the resonant features for the particular structure we study in this paper, the understanding gained from Eqs. (5-8) can help in the design of future studies. For example, replacing the highly polarizable incipient ferroelectric SrTiO$_3$ with a different insulator with smaller dielectric response will make the transfer of charge energetically unfavorable. A possible candidate is SrZrO$_3$ which is isovalent to SrTiO$_3$ but has much smaller lattice polarizability.

Another interesting bit of information gleaned from Eq. (8) is that the charge transfer depends on the total thickness of the barrier. So while a given material combination may exhibit pure tunneling behavior for thin barriers, as we found for TJ$_2$, as the barrier thickness increases there will be an increase in the population of the 2DEG due to the growing electrostatic energy cost given in Eq (5). This opens the possibility of observing a *quantum* transition to resonant tunneling as the Fermi surface of the 2DEG expands and the polar distortion decreases, eventually acquiring states with orbital character that contribute significantly to tunneling as for TJ$_1$. Such a transition will show up in an experiment as an *increase* in conductance as thickness increases.

## V. SUMMARY

We have performed first-principles electronic structure and transport calculations of all oxide heterostructure tunnel junctions supporting a 2DEG within the barrier. This introduces the possibility of vertical connectivity to complement the horizontal in-plane architectures already proposed, such as all-oxide high-mobility field-effect transistors. The particular resonant tunnel junction considered involves a SrTiO$_3$/LaO/SrTiO$_3$ barrier placed between two SrRuO$_3$ electrodes. We show that the tunneling through such a junction can be strongly resonant because the 2DEG trapped in the middle of the barrier acts as an extended 2D potential well, essentially making the system a double barrier tunnel junction. Our calculations indicate that the 2DEG formed in the barrier enhances the conductance by a factor of $10^5$. In addition, this resonant tunneling is strongly selective of the orbital symmetry of the tunneling states due to the electronic structure of the 2DEG. However, we also find that there is a competition between populating the 2DEG and polarizing the surrounding dielectric in order to screen the (LaO)$^+$ layer. Replacing the SrTiO$_3$ with other materials with smaller polarizability or tuning the device geometry could lead to feasible device structures. This may prove to be an interesting component in the design of all oxide electronics.

## ACKNOWLEDGMENTS

This work is supported by NSF-MRSEC, the Nanoelectronics Research Initiative and the Nebraska Research Initiative. Work at UPR was supported by IFN (NSF Grant No. 0701525). Computations were performed utilizing the Research Computing Facility of the University of Nebraska–Lincoln and the Center for Nanophase Materials Sciences (CNMS) at Oak Ridge National Laboratory.* Electronic address: jdburton1@gmail.com